\documentstyle[epsf]{mn}

\newcommand{\mnras}[1]{MNRAS}
\newcommand{\apj}[1]{ApJ}
\newcommand{\apjs}[1]{ApJS}
\newcommand{\apjl}[1]{ApJL}
\newcommand{\nat}[1]{Nature}
\newcommand{\aap}[1]{A\&A}
\newcommand{\araa}[1]{ARA\&A}
\newcommand{\aaps}[1]{A\&ASS}
\newcommand{\aj}[1]{AJ}
\newcommand{\apss}[1]{Ap\&SS}

\title[]{2016+112: A Gravitationally Lensed Type--II Quasar}
\author[Koopmans et al.]{L.V.E.\,Koopmans$^1$, M.A. Garrett$^2$,
  R.D. Blandford$^1$, C.R. Lawrence$^3$, \and A.R. Patnaik$^4$ \&
  R.W. Porcas$^4$ \\ $^1$California Institute of Technology, mailcode
  130-33, Pasadena CA 91125, USA \\ $^2$Joint Institute for VLBI in
  Europe, P.O.Box 2, NL--7990 AA, Dwingeloo, The Netherlands \\
  $^3$Jet Propulsion Laboratory, California Institute of Technology,
  4800 Oak Grove Drive, Pasadena, CA 91109, USA \\
  $^4$Max-Planck-Institut f\"{u}r Radioastronomie, Auf dem H\"{u}gel
  69, D-53121 Bonn, Germany}
\date{Accepted ... Received ...}
\pubyear{2001}

\begin{document}

\maketitle

\begin{abstract}
A single-screen model of the gravitational lens system 2016+112 is
proposed, that explains recent {\sl Hubble Space Telescope} (HST)
infrared (NICMOS--{\sl F160W}) observations and new high-resolution
{\sl European VLBI Network} (EVN) 5--GHz radio observations, presented
in this paper. In particular, we find that a massive `dark' structure
at the lens position, previously suggested by X-ray, optical and
spectroscopic observations of the field around 2016+112, is not
necessarily required to accommodate the strong lensing constraints. A
massive structure to the north-west of the lens system, suggested from
a weak-lensing analysis of the field, is included in the model. The
lensed source is an X-ray bright active galaxy at $z$=3.273 with a
central bright optical continuum core and strong narrow emission
lines, suggestive of a type--II quasar. The EVN 5--GHz radio maps show
a radio-jet structure with at least two compact subcomponents. We
propose that the diamond caustic crosses the {\sl counter-jet} of the
radio source, so that part of the counter-jet, host galaxy and
narrow-line emission regions are quadruply imaged. The remainder of
the radio source, including the core, is doubly imaged.  Our lens
model predicts a very high magnification ($\mu$$\sim$300) at the
bightness peaks of the inner two radio components of complex~C. If the
jet exhibits relativistic velocities on micro-arsecond scales, it
might result in apparent {\sl hyperluminal} motion. However, the lack
of strong radio variability and the peaked radio spectrum imply that
these motions need not be present in the source.  Our model
furthermore implies that the optical spectrum of C$'$ can
only show features of the AGN and its host galaxy.
\end{abstract}


\section{Introduction}

The gravitational lens system 2016+112, discovered by Lawrence et al.
(1984), has defied any simple explanation. The system consists of two
AGN images (A and B) at a redshift of $z$=3.273 (Lawrence et al. 1984;
Schneider et al. 1985, 1986), separated by 3.4 arcsec. Early optical
and near-infrared observations (e.g. Schneider et al. 1985; Langston,
Fischer \& Aspin 1991; Lawrence, Neugebauer \& Matthews 1993) showed
the presence of two extended objects (designated C$'$\footnote{The
prime distinguishes it from the radio structure (designated~C), which
early on was not known or thought to be associated with the optical
emission.} and D). C$'$ lies some 2 arcsec south-east of image B, such
that ABC$'$ nearly form a right triangle. Object D, the primary lens
galaxy, has a redshift of 1.01 (Schneider et al. 1985) and is close to
the centroid of A, B and C$'$. The redshift of C$'$ has resisted
measurement.  Narrow Ly-$\alpha$ emission near A, B and C$'$ has been
detected at a redshift of $z$=3.273 with line widths
$\la$1000~km\,s$^{-1}$ (Schneider et al. 1986, 1987; Lawrence 1996;
Yamada et al. 2001; Lawrence et al. 2002).  Similarly, two fuzzy
patches of Ly-$\alpha$ emission were detected about 3 arcsec
north-west and west of images A and B, respectively (Schneider et
al. 1986, 1987). Recently, high-resolution {\sl F814W}--band (Lawrence et
al. 2001) and {\sl F160W}--band observations were obtained with the HST (see
the CASTLES webpage http://cfa-www.harvard.edu/glensdata/MG2016.html;
e.g. Mu\~noz et al. 1998), showing that images A and B are unresolved
and compact, whereas C$'$ is arc-like and does not show obvious
compact structure.

At radio wavelengths, 2016+112 has been observed with the VLA
(e.g. Lawrence et al. 1984; Schneider et al. 1985) and displays three
components, A, B and C. All components contain compact substructure in
higher resolution MERLIN (Garrett et al. 1994), VLBI (Heflin et
al. 1991) and EVN (Garrett et al. 1996) images. Whereas components A
and B consist of at least two subcomponents (oriented approximately
north-west), radio component C splits into four dominant subcomponents
aligned east-west (e.g. Garrett et al. 1996). Components A and B have
steep integrated radio spectra between 18\,cm and 6\,cm (Lawrence et
al. 1994), whereas component C has a somewhat flatter integrated radio
spectrum. All component spectra steepen between 6 and 2\,cm (Langston
et al. 1991). The overall integrated spectrum of the lens system is
that of a GPS source and peaks somewhere in the range of 1 to 5
GHz. With the VLA in A-array, the source is barely detectable at
22~GHz (Patnaik, private communication). Observations with EVN 
at 18\,cm and MERLIN at 6\,cm (Garrett et al. 1996) show that the outermost
images (C$_{11}$ and C$_2$) have significantly steeper spectra than
the two innermost components (C$_{12}$ and C$_{13}$). These
observations suggest that the structures C$_{11}$+C$_{12}$ and
C$_{13}$+C$_{2}$ have opposite parities and are probably images of the
same structure in the source.

Observations of the field around 2016+112 with the {\sl ASCA satellite
X-ray Observatory} suggested the presence of diffuse X-ray emission
centered on the lens system and to the north-west (Hattori et
al. 1997; Benitez et al. 1999). Recent observations with the {\sl
Chandra X-ray Observatory}, however, unambiguously show that this
emission is mostly due to discrete sources and not diffuse cluster
emission (Chartas et al. 2001). In addition, the lensed images A and
B, and complex C$'$, are detected. The presence of X-ray and narrow
emission lines suggests that the source could in fact be a type--II
quasar (e.g.  Yamada et al. 2001; Chartas et al. 2001). None of the
other X-ray sources in the field are associated with galaxies detected
by Soucail et al. (2001; see below). The comparatively large number of
X-ray sources in the field of 2016+112 might be the result of a
magnification bias, which enhances the number count of sources around
the lensing mass distribution if the number density of X-ray source
increases steeply at redshifts larger than that of the lens galaxy.

Spectroscopic observations by Soucail et al. (2001) show the presence
of an overdensity of galaxies at the same redshift as lens galaxy~D.
Similarly, Clowe et al. (2001) detect a 3--$\sigma$ weak-lensing
signal to the north-west of the lens system, roughly coincident with
the region of excess X-ray emission (e.g. Benitez et al. 1999).  The
latter could again be due to a magnification bias.  Only a marginal
weak-lensing signal at the position of the lens system was found,
consistent with the absence of a very massive dark X-ray cluster.  If
the dispersion in velocities of the field galaxies, found by Soucail
et al. (2001), are representative of a virialized system, one would
have expected to see diffuse X-ray emission (Chartas et
al. 2001). Absence of the latter therefore also suport that these 
galaxies have not yet virialized and formed a massive centrally
concentrated cluster.

Several models have been proposed to explain these observations. Some
employ a single deflector and a single-screen (Langston et al. 1991;
Benitez et al. 1999), whereas others used more complex models with
C$'$ and D being different galaxies (Narashima et al. 1987). In the
proposed two-screen models it is assumed that object C$'$ is a galaxy
at a redshift different from galaxy D, and responsible for lensing
radio complex C into additional `subimages' (Nair 1993; Nair \&
Garrett 1997). The mirror-symmetry and opposite parities for the
structures C$_{11}$+C$_{12}$ and C$_{13}$+C$_{2}$ suggest that the
source structure corresponding to complex C is quadruply imaged and
that there is no need to invoke a second lens screen, that is not
a weakly perturbative, but changes the nesting of the caustic curves
in the sources plane. However, a
two-screen model or a model that has two lens galaxies in the same lens
plane, predicting the same parities as a single-screen single-lens model
(Nair \& Garrett 1997) can of course not be excluded based on this argument
alone.

In this paper we propose an alternative model that can quantitatively explain these
observations with a single screen. Instead of explaining the lens
system with a complex deflector model, we find it can also be explained by
a realistic, although more complex source model. In Sect.2, we present
EVN 5--GHz radio observations of 2016+112, which suggest that
complex C consists of two images with opposite parities.  In Sect.3, a
detailed structure of the source is proposed which qualitatively can
explain most of the observed features of the 2016+112 lens system. In
Sect.4, we present a model of the lens potential that incorporates
mass structures from the field around the lens system. In Sect.5, we
compare the model to the observational constraints to find that it can
also quantitatively explain the available observational constraints of
the lens system and the field. In Sect.6, our results are
summarized and discussed. Throughout this paper we assume a flat
smooth Friedman-Robertson-Walker (FRW) universe with $\Omega_{\rm
m}$=0.3 and H$_0$=65~km\,s$^{-1}$\,Mpc$^{-1}$.

\begin{figure*}
\begin{center}
\leavevmode
\hbox{%
\epsfxsize=\hsize
\epsffile{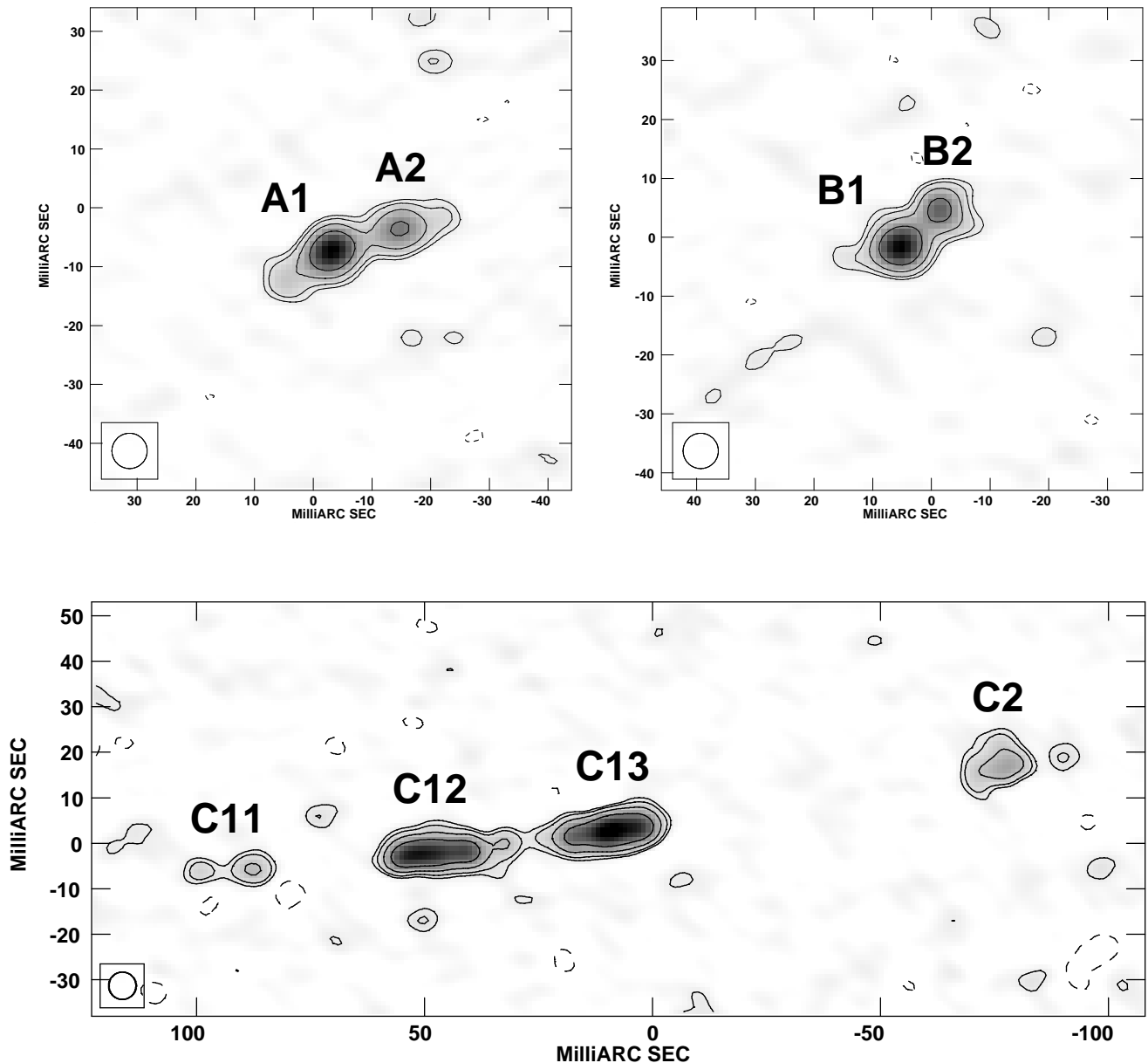}}
\end{center}
\caption{{\sl European VLBI Network} (EVN) 5--GHz radio images of 
regions A, B and C of 2016+112. Contour levels indicate $(-3, 3, 5,
10, 20)$ times the rms noise level of 0.14 mJy\,beam$^{-1}$.  One
notices the jet-like substructure in images A and B, and the apparent
`mirror-symmetry' of components C$_{12}$ and C$_{13}$, suggesting
opposite parties due to lensing. The model proposed in this paper
assumes that both components are images of a quadruply-lensed part of
the source, which is associated with the counter-jet in components
A$_2$ and B$_2$, the core of the source. Components C$_{11}$ and
C$_{2}$ are presumably steep-spectrum images of intrinsically faint
(but highly magnified) substructure further along the counter-jet
(i.e. west of A$_2$ and B$_2$). The restoring beam has a FWHM of 6
milli-arcsec.}
\end{figure*}

\section{EVN 5--GHz Observations and Data Analysis}

The 5--GHz EVN observations were made on 17--18 May 1995 from UT 22h30 to
10h30, using 7 antennas of the EVN: the 100-m Effelsberg (DE), 26-m
Jodrell-Mk2 (UK), 25-m Onsala (SE), 32-m Medicina (IT), 32-m Noto (IT),
32-m Cambridge (UK), and Westerbork-array (NL).  Technical problems
resulted in no data from the latter two antennas. The recording mode
was MkIIIa Mode A (56\,MHz bandwidth, Left Hand Circular polarisation).
A phase-reference observing scheme was used, switching between the
target, 2016+112, and a compact calibrator, J2029+121, located 1 degree
away. The tape ran continuously through each 13 minute pass, and the
source switching cycle consisted of alternating between 90 seconds on
the calibrator and 140 seconds on the target. Correlation of the data
was conducted at the MPIfR MkIIIa correlator in Bonn. To prevent loss
of data during correlator synchronisation of the very short scans, each
baseline was correlated twice as a continuous tape pass, once at each
of the two source positions. (Spurious correlations on the "wrong"
source were edited out in the later data analysis).              

Subsequent data processing and analysis was performed with the 
NRAO~{\sc AIPS} package. Long fringe-fitting solutions were made that
included several switching phases of the calibrator; the resulting
phase, delay and rate solutions were then applied to 2016+112.
Although the interferometer model used in the MPIfR correlator was not
sufficiently accurate to permit direct phase-referencing, this process
did remove short-term phase fluctuations, and allowed longer solution
intervals to be used for self-calibration of the phases of the
2016+112 visibilities. Amplitude calibration was made by initially
assuming that J2029+121 is a 0.91--Jy point source on all baselines,
and then determining self-calibration corrections after mapping the
source.  These amplitude and phase corrections were also applied to
2016+112. In order to prevent fringe-rate and delay smearing over the
4 arcsecond field of view, the data associated with 2016+112 were
maintained as 28 contiguous 2~MHz channels with a visibility averaging
time of 2.5 seconds. The calibrated 2016+112 data were Fourier
transformed and a naturally weighted, tapered image of the full field
was produced. All three main regions of emission A, B and
C were clearly detected and the image was used as an input model
for subsequent (phase-only) self-calibration.  Since 2016+112 is a
rather faint, resolved radio source, improvements to the original
phase solutions were only obtained by employing a relatively long
solution interval (13 minutes) over the entire frequency band (56
MHz). CLEANed maps of the three main regions of emission are presented
in Fig.1. All the maps are naturally weighted (the rms noise level is
$\sim$140\,$\mu$Jy beam$^{-1}$) and the FWHM of the circular restoring
beam is 6~mas. 

To obtain the positions and flux-densities of the radio components in
regions A, B and C, elliptical Gaussians were fitted to the
images. Images A and B were fitted by two Gaussians. Images
C$_{12}$ and C$_{13}$ could not be fitted by single elliptical
Gaussians and were therefore each fitted by two Gaussians ($a$ and $b$,
respectively). The results of these fits are listed in Table~1.

\begin{table*}
\centering
\begin{tabular}{lllccc}
\hline
Comp. & RA (arcsec) & Dec (arcsec) & S$_{\rm 5\,GHz}$ (mJy) & PA ($^\circ$) & p$_\pm$ \\
\hline
A$_{1}$   & $+$0.0000 & $+$0.0000 & 8.0(0.3)   & $-$71.7   & $+$\\
A$_{2}$   & $-$0.0121 & $+$0.0040 & 4.5(0.3)   & --        & $+$\\
B$_{1}$   & $-$3.0057 & $-$1.5040 & 7.3(0.3)   & $-$48.5   & $-$\\
B$_{2}$   & $-$3.0126 & $-$1.4979 & 4.2(0.3)   & --        & $-$\\
C$_{11}$  & $-$2.0111 & $-$3.2331 & 1.9(0.3)   & --        & $-$\\
C$_{12a}$ & $-$2.0455 & $-$3.2300 & 3.8(0.3)   & $+$95.1   & $-$\\
C$_{12b}$ & $-$2.0542 & $-$3.2294 & 13.9(0.5)  & $+$93.9   & $-$\\
C$_{13a}$ & $-$2.0937 & $-$3.2241 & 8.7(0.4)   & $-$80.2   & $+$\\
C$_{13b}$ & $-$2.0855 & $-$3.2254 & 9.8(0.4)   & $-$81.0   & $+$\\
C$_{2}$   & $-$2.1749 & $-$3.2101 & 5.6(0.5)   & --        & $+$\\
\hline
\end{tabular}
\caption{Properties of the image components determined from the 5--GHz
EVN data (Sect.2).  Column (1): The component's name as in Garrett et
al. (1996). Columns (2-3): The positions determined from Gaussian fits
to the 5--GHz EVN images. The positional error is 1~mas with respect
to A$_{1}$.  Column (4): The total flux-density and error on the last
significant digit (within parentheses) of each of the
components. Column (5): The position angles (PA) of the presumed jet
direction, measured in the direction A$_{1}$$\rightarrow$A$_{2}$,
B$_1$$\rightarrow$B$_2$,
C$_{12b}$$\rightarrow$C$_{12a}$$\rightarrow$$C_{11}$ and
C$_{13b}$$\rightarrow$C$_{13a}$$\rightarrow$$C_{2}$ from north to
east. Column (6): The parities p$_\pm$, based on general assumption of
the source structure and geometry of the time-delay surface
(e.g. Blandford \& Narayan 1986).}
\end{table*}

\section{The Source}

Based on the optical/IR and radio observations of 2016+112 (see
Sections 1 and 2), we postulate the following simple picture for the
lensed source (see also Langston et al. 1991 and Benitez et al. 1999
for analogous models). First, we assume that optical continuum
emission of images A and B is from an AGN (possibly a type--II quasar)
near the core of some faint underlying host galaxy, which is not
detected at A and B due to its very low surface brightness compared to
the AGN.  Second, the NW elongation of the EVN 1.7~GHz radio images of
A and B (Garrett et al. 1996), is now confirmed by the high-resolution
EVN observations at 5~GHz presented in Sect.2, showing that both
components consist of at least two subcomponents.  We assume that the
flatter-spectrum subcomponents seen at 5~GHz (A$_2$ and B$_2$) are
associated with the optical continuum core and the other two (A$_1$
and B$_1$) are jet-features. Because the constraints from the optical
emission are not used at this level of positional accuracy (few mas),
this particular choice is of little relevance for the lens modeling.

\begin{figure}
\begin{center}
\leavevmode
\hbox{%
\epsfxsize=\hsize
\epsffile{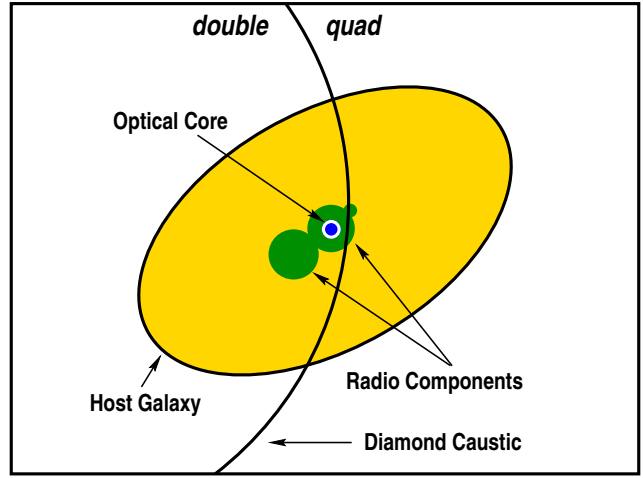}}
\end{center}
\caption{A schematic view of the proposed source structure for
2016+112. The AGN, associated radio core, part of host galaxy and the
second radio component are doubly imaged. The western-most part of the
radio core and part of the host galaxy are quadruply imaged. The
components are not drawn to scale, nor do the ellipticity and position
angles of host galaxy and other structures necessarily represent those
of the true underlying source.}
\end{figure}

Additionally, strong spectral lines (e.g. Ly--$\alpha$,
C--{\footnotesize IV}, N--{\footnotesize V}, Si--{\footnotesize IV})
are seen near images A, B and C$'$ (e.g. Schneider et al. 1986, 1987;
Lawrence 1996; Yamada et al. 2001). However, the line widths are very
narrow compared to a typical quasar spectrum drawn from the same
redshift range (see Steidel et al. 2001). This is a puzzle, although
it could be that line emission from the BLR is obscured, as suggested
by Yamada et al. (2001), and that only the NLR further from the core
is seen. This would also explain why the Ly-$\alpha$ emission is not
exactly coincident with A, B and C$'$, but offset from C$'$ by 1 WFPC2
pixel (0.0455$''$) when aligned with A and B. These narrow-line spectra are
indicative of a type--II quasar (Yamada et al. 2001), which is
supported by the X-ray brightness of the source (Chartas et
al. 2001). A schematic picture of the source is shown in Fig.2.

Qualitatively, this source structure can explain both the optical and
radio data of 2016+112 available to date, if the diamond caustic
(fold) crosses the source as indicated in Fig.2: (i) the optical core
is doubly imaged at A and B, because it falls outside the diamond
caustic, (ii) most of the associated radio structure is also doubly
imaged near A and B, (iii) only part of the radio structure,
associated with the counter-jet near radio subcomponents A$_2$ and
B$_2$, is quadruply imaged, such that C$_{12}$ and C$_{13}$ are
fainter than what one might expect based on their proximity to the
critical curve (i.e. the high magnification) and the flux-density of
A$_2$ and B$_2$, (iv) components C$_{11}$ and C$_2$ are steep-spectrum
emission further along the counter-jet, which are quadruply imaged but
only detected near C because of the very high magnification near the
critical curve, (v) part of the underlying host galaxy is doubly
imaged, but is too faint to be seen at images A and B as a result of
the bright AGN emission (and its associated PSF), (vi) the other part
of the host galaxy is quadruply imaged, resulting in the highly
magnified arc (i.e. complex~C$'$), (vii) similarly the extended
Ly-$\alpha$ emission (Lawrence et al. 2002), as well as the
narrow-line emission in general (e.g. Yamada et al. 2001) is expected
to surround the optical core and is part doubly and part quadruply
imaged (i.e. that part inside the diamond caustic) and therefore seen
near A, B and C$'$. This is supported by the idea that the
emission-line ratios near C$'$ are more consistent with it originating
further from the core (Yamada et al. 2001).  Although the proposed
model might appear very complicated, it is precisely the complex
lensed structure one expects from AGNs -- which are wealthy in
wavelength-dependent structure on sub-arcsec scales -- if they happen
to cross a caustic.

Support for the suggestion that part of the radio source crosses the
caustic is given by the actual `merger' of radio components C$_{12}$
and C$_{13}$ at the few--$\sigma$ level of the surface-brightness
contours in Fig.1. This {\sl only} occurs if part of the source at
that surface-brightness level actually crosses the caustic in the
source plane. The critical curve crosses precisely that point in the
image plane where the two images merge (i.e. the `saddle-point' in the
surface brightness distribution between C$_{12}$ and C$_{13}$; see
e.g. Kochanek, Keeton \& McCleod 2000). Due to the conservation of
surface brightness in gravitational lensing, we expect the point where
images C$_{12}$ and C$_{13}$ merge to be accociated with extended
structure north-west of A$_2$ and B$_2$ at the few--$\sigma$ contour
level as well (see Fig.1). Images C$_{12}$ and C$_{13}$ are associated
with very compact (i.e. a few $\mu$as) substructure in the
counter-jet, which unfortunately is barely observable near images A
and B, due to their relatively low magnifications (see Sect.5 below)
and of course the finite resolution of the observations. This is even
more strongly the case for the faint and steep-spectrum emission from
C$_{11}$ and C$_2$.

In the model by Langston et al. (1991) the second flat-spectrum source
component (i.e. the component associated with A$_2$ and B$_2$) lies
almost fully {\sl inside} the diamond caustic, whereas the steeper
spectrum component straddles the caustic. Nair \& Garrett (1997)
showed, however, that this particular model predicts the inner two
images of complex C to be steep spectrum and the outer two flat
spectrum, opposite to observations.  On this basis, the model from
Langston et al. (1991) was rejected.  In the model proposed here,
however, the steeper spectrum component lies fully {\sl outside} the
diamond caustic, whereas only a small part of the flat-spectrum
component straddles the caustic from the outside. This small part
(i.e. the counter-jet) inside the caustic is quadruply imaged and
highly magnified, creating images C$_{12}$ and C$_{13}$. Note that the
direction of merging images C$_{12}$ and C$_{13}$ is perpendicular
to the caustic near the source position (Fig.3 below), as must
generally be true for a single screen model (Blandford \& Narayan 1986). 
Further inside the
caustic, we postulate that the spectrum of the counter-jet steepens
and that a small subcomponent is lensed in to images C$_{11}$ and
C$_2$. Both of these are only seen near C, because of their
exceedingly high magnifications (Sect.5).

This type of lens--source configuration is not uncommon. Several radio
gravitational lens systems have been observed with part of the source
inside and part of the source outside the diamond caustic
(e.g. Einstein rings). In particular, the JVAS/CLASS lens B1938+666
has a radio jet structure crossing the diamond caustic near the cusp,
although in this case most of the source lies inside the caustic (King
et al. 1997), wheres in 2016+112 only the counter-jet is quadruply
imaged.

The question is now whether we can also {\sl quantitatively} explain
the {\sl primary constraints} (e.g. radio and optical image positions,
the jet position angles (PA) and the flux-density ratios), and whether the
resulting model is in agreement with {\sl secondary constraints}, such
as the properties of the host galaxy (D) and the detection of one or
more nearby mass concentrations in weak-lensing (Clowe, Trentham 
\& Tonry 2001) and spectroscopic studies (Soucail et al. 2001).

\begin{table*}
\centering
\begin{tabular}{ccccccc}
\hline
Defl. & RA ($''$)  & Dec ($''$)& $f$ & PA ($^\circ$)& $\sigma_{||}$ (km\,s$^{-1}$)\\
\hline
D & [$-$1.740] &  [$-$1.782] & 0.75--0.77 & [$-$59] & 320$\rightarrow$342 \\
M$_1$ & [$-$24.0] & [+60.0] & [1.0] & [0.0] & [970]\\
M$_2$ & 11.6$\rightarrow$2.0 & 0.8$\rightarrow$$-$0.4  
	& [1.0] & [0.0] & [560]$\rightarrow$[175]\\
\hline
\end{tabular}
\caption{The fixed (between brackets) and reconstructed mass-model
parameters for the primary lens galaxy D and the two SIS mass
distributions M$_1$ and M$_2$. The arrows indicate how the values
change as the velocity dispersion of M$_2$ is lowered from 560 to 175
km\,s$^{-1}$. In addition to the mass distributions, an external shear
($\vec{\gamma}_{\rm ext}$) is found with strength
0.12$\rightarrow$0.07 and position angle
$-$24$^\circ$$\rightarrow$$-$51$^\circ$.}
\end{table*}

\section{The Lens \& Field}

First, we associate a singular isothermal elliptical (SIE) mass
distribution (e.g. Kassiola \& Kovner 1993; Kormann, Schneider \&
Bartelmann 1994; Keeton, Kochanek \& Seljak 1997) with the primary
lens galaxy D, although a mass model with a steeper or more shallow
mass profile is not a priori excluded (e.g. Benitez et al. 1999).  In
light of the present uncertainties, especially about external
perturbers, we feel it is not yet warranted to explore more detailed
lens mass models. We associate the center of the mass distribution
(MD) with the centroid of the surface brightness distribution (SBD) of
galaxy D, measured from the {\sl Hubble Space Telescope} (HST) NICMOS
{\sl F160W}--band image (CASTLES; e.g. Mu\~noz et al. 1998),
i.e. ($-$1.740$\pm$0.003, $-$1.782$\pm$0.003) arcsec with respect to
image~A. In general, good agreement is found between position angles
of the SBD and MD of lens galaxies (Keeton, Kochanek \& Seljak
1997). To avoid underconstraining the mass model, we have chosen to
fix the position angle of the MD at the observed SB value of
$-59$$\pm$2 degrees (see CASTLES web page). The only free parameters
of the MD of lens galaxy D are therefore the axial ratio ($f$) and the
central velocity dispersion ($\sigma_{||}$; as defined in Kormann et
al. 1994).

To model the surrounding field, we place a singular isothermal sphere
(SIS) mass distribution (designated M$_1$) at the position ($-$24.0,
$+$60.0) arcsec NW of image~A, where Clowe et al. (2001; see
Sect.1) find a significant weak-lensing signal. A SIS fit to the weak-lensing
shear field suggests a velocity dispersion of 970$^{+150}_{-180}$
km\,s$^{-1}$ (1--$\sigma$ errors; assuming a redshift of unity). We
constrain the velocity dispersion of M$_1$ to this value. In addition,
Soucail et al. (2001) have spectroscopically confirmed the presence of
an overdensity of at least six  in red galaxies at the redshift of lens
galaxy~D. They estimate a velocity dispersion of
771$^{+430}_{-160}$~km\,s$^{-1}$. Clowe et al. (2001) do not to find a
strong shear signal at this position ($\sigma$=560$^{+220}_{-480}$
km\,s$^{-1}$; 1--$\sigma$ errors). The low SNR of the observations,
however, does not allow one to exclude it. We include this mass
distribution to first order by modeling it as a SIS (M$_2$). This is
done in order to test whether an additional mass distribution is
indeed required by the strong-lensing constraints, as suggested by
these previous authors. We allow the velocity dispersion of M$_2$ to
vary between 560--175~km\,s$^{-1}$ and let its position be free. The
upper limit is roughly defined, such that M$_2$ does not result in
additional observable images of the background source, although this
constraint could be lifted if this mass distribution has a large
finite core and is not capable of multiple imaging.  Similarly, the
lower limit avoids additional sub-images of A, when M$_2$ nearly
coincides with its position (see Sect.5). In the latter case, the
velocity dispersion of M$_2$ is of galactic scale (few
hundred~km\,s$^{-1}$), in which case the core radius is typically
small.  We include an external shear with both its strength and
positional angle as free parameters.

\subsection{Constraints}

As primary constraints on this starting model, we use the observed
properties of images A, B, C$_{12}$ and C$_{13}$, as found from the
HST NICMOS {\sl F160W}--band (e.g. CASTLES) and EVN 5--GHz images
(Sect.2; Table 1).  The relative positions of A and B in the optical
and radio agree to within the measurement errors. There appears to be
a slight offset in declination between the {\sl F160W}--band emission
of complex C$'$ in the HST image and the positions of the EVN radio
components. It amounts, however, to only 1 pixel (0.1$''$) and in
light of the uncertain structure of the host galaxy, we consider this
difference unimportant at present. The optical brightness ratio between images
A and B, found from the HST NICMOS {\sl F160W}--band image is 
$r_{\rm n}$=0.97$\pm$0.02 (i.e. $S_{\rm B}/S_{\rm A}$). Garrett et
al. (1996), however, find $r_{\rm n}$=0.84$\pm$0.01 at 1.7 GHz. We
therefore adopt an ``average'' value $r_{\rm n}$=0.90$\pm$0.10 for
the flux-ratio between images A and B, in our lens modeling,
where the uncertainty accounts for possible variability,
even though the
source does not appear to vary strongly (Haarsma et al. 2001). 
The adopted value is also consistent with $r_{\rm n}$=0.92$\pm$0.05 found 
from the EVN 6-cm observations, presented in this paper (Table~1). 

We do not use the positions of images C$_{11}$ and C$_2$ as
constraints.  According to the source model (Sect.3; Fig.2), they are
part of the source structure associated with lensed images A$_2$ and
B$_2$. They appear somewhat more resolved in the EVN 5--GHz
observations (Sect.2), although the signal-to-noise is low, and have
steeper spectra compared with the other images between the EVN 1.7 and
5--GHz. Hence, they are probably not part of the same region of the
source that is responsible for images $A_2$, B$_2$, C$_{12a/b}$ and
C$_{13a/b}$ seen in Fig.1. The emission from C$_{11}$ and C$_2$ most
likely originates further along the counter-jet, which presumably has
a steeper spectrum (Sect.3). We assume that the emission from
C$_{12a}$ and C$_{13a}$ is associated with structure of the unresolved
images A$_2$ and B$_2$, within a 1-mas radius from their respective
centroids.  Although this particular choice might seem arbitrary, it
is probably conservative based on the notion that from an inverse
Compton limit on the brightness temperature of these radio sources,
between 10$^{11}$ and 10$^{12}$~K, a component size for A$_2$ and
B$_2$ as small as 0.1--0.3~mas can be expected, given their observed
flux-densities. If the axial ratios of the components are large
(i.e. jet-like), the emission region could easily `stretch' to 1~mas
or larger in one direction.  We do {\sl not} use the flux-density
ratios between images A$_2$, B$_2$ and complex C, because the source
structure lies on a caustic and consequently has a strong
magnification gradient over its extent.  The source can therefore not
be treated as a point source, rendering the use of a flux-density
ratio very difficult.  All constraints are listed in Table~1.

\begin{figure*}
\begin{center}
\leavevmode
\hbox{%
\epsfxsize=0.87\hsize
\epsffile{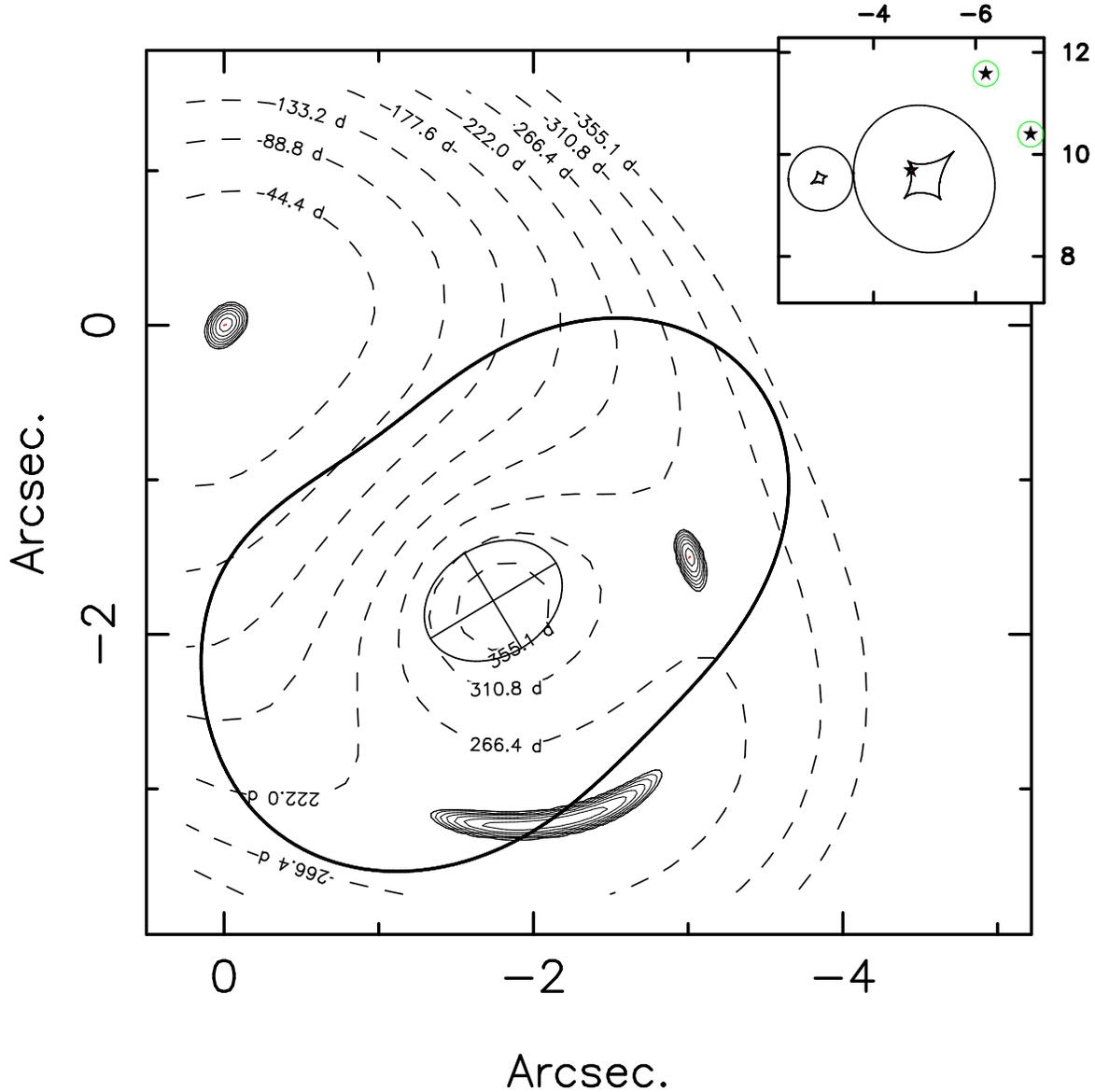}}
\end{center}
\caption{The model of 2016+112 as indicated in Table 2, assuming M$_2$ is
an $L_*$ galaxy with $\sigma_{||}$=225~km\,s$^{-1}$ at
($-$3.2$''$,$-$0.4$''$) from image~A. The thick line indicates the
critical curve. The ellipse indicates the position, flattening and
position angle of the primary lens galaxy (i.e. D), whereas the dashed
lines indicate constant time-delay contours. The subpanel (units in
arcsec) shows the caustics of galaxies D and M$_2$, the source
(indicated by the star inside the caustics) and the two Ly--$\alpha$
patches (stars with circles; see Sect.6). The caustics are displaced
to the north from the primary lens galaxy by $\sim$10 arcsec due
to a massive structure M$_1$, inferred from weak lensing (see text).}
\end{figure*}

\section{Results}

Using the mass model and constraints discussed above, we vary the nine
free parameters using the simulated annealing downhill simplex method
described in Press et al. (1992), until the difference between the
observed and recovered image properties are minimized in terms of the
$\chi^2$--value (e.g. goodness-of-fit). Using different starting
values of the parameters, we ensure that the final solution is close
to the absolute minimum in the $\chi^2$--space.

We vary the velocity dispersion of M$_2$ between 560 and
175~km\,s$^{-1}$ (see above) and minimize $\chi^2$ for the other free
parameters.  The goodness-of-fit $\chi^2$ (for two degrees of freedom)
of the best models increases only marginally between these upper and
lower limits on the velocity dispersion of M$_2$, i.e. from 1.1 to 1.5
respectively.  This indicates a strong degeneracy in the lens model,
between the velocity dispersion and position of M$_2$ and the external
shear. The velocity dispersion and axial ratio of G1 are only
marginally affected. We note that the $\chi^2$--values can not be used
to calculate a likelihood for the model, because we did not strictly
use measurement errors for all constraints. The image positions,
flux-ratio, PA's and parities are recovered in good agreement with the
constraints (Table~1), given the uncertainties in the structure of the
lensed images.  The recovered image properties, inferred
magnifications and time-delays are listed in Table~3.  In Fig.3, we
show the image configuration, critical curve, caustics and time-delay
surface of this model.

In Table~2, we have listed the recovered parameters of the lens galaxy
D and two additional mass structures M$_1$ and M$_2$.  We find that
the SD axial ratio $f$=0.75--0.76 of galaxy D is somewhat larger than
that determined from its SB distribution, i.e. $0.57\pm0.01$, as seen
in the HST NICMOS--{\sl F160W} observations, (e.g. CASTLES web page;
see also Benitez et al.  1999). The velocity dispersion of
320--340~km\,s$^{-1}$ of galaxy D implies a rest-frame mass-to-light
ratio of $(M/L)_{\rm H}$$\approx$1.7$\cdot h_{65}$\,M$_\odot$/L$_{\rm
H, \odot}$ (no evolutionary correction), assuming a singular
isothermal sphere to calculate the mass inside the Einstein radius, an
{\sl F160W}--band magnitude for lens galaxy D of 18.12$\pm$0.04 (see
CASTLES web-page) and the galaxy models from Poggianti (1997). This
mass-to-light ratio compares well with those found for other lens
galaxies (e.g. Jackson et al. 1998), which indicates that the image
separation should not be significantly affected by a mass contribution
with a different mass-to-light ratio (e.g. a dark cluster). In that
case, the lens galaxy would most likely not have been on the
Fundamental Plane of early-type galaxies either (Kochanek et
al. 2000).

The model also indicates the presence of an external shear with a
strength $\gamma$ between 0.07 and 0.12 and a position angle between
$-$24 and $-$51 degrees, depending on the velocity dispersion of M$_2$.  
If we investigate the {\sl I}--band image
(Fig.2 in Soucail et al. 2001) in more detail, there appears to be a
`filamentary structure' of high-redshift galaxies running across the
lens system. This `filament' is {\sl not} dynamically related, because 
it contains galaxies over a very different range of redshifts ($z\approx$0.6--1.1), 
but it might be responsible for the shear at the position
of 2016+112. On the other hand, the closeness of the position angle of the
external shear and that of the SB of galaxy D, for low velocity dispersions of
M$_2$, might also indicate that either the mass distribution of galaxy D has
a different radial mass profile or that its flattening is a function of radius.

\begin{table*}
\centering
\begin{tabular}{lccccc}
\hline
Comp. & RA (arcsec) & Dec (arcsec) & $r_{\rm \{B,C\}/A}$ & $\mu$ & $\Delta t$/$h_{65}$ (days)\\
\hline
A$_{1}$   & $+$0.0008/0.0008 & $-$0.0002/0.0000 & [1.0]   & $+$4.0/3.3 & $\equiv$0.0\\
A$_{2}$   & $-$0.0121/0.0121 & $+$0.0040/0.0040 & [1.0]   & $+$4.0/3.3 &  3.3/9.1\\
B$_{1}$   & $-$3.0060/3.0060 & $-$1.5036/1.5033 & 0.91/0.85    & $-$3.6/2.8 &  255.3/291.6\\
B$_{2}$   & $-$3.0126/3.0127 & $-$1.4979/1.4981 & 0.90/0.84    & $-$3.6/2.8 &  254.9/286.4\\
C$_{12b}$ & $-$2.0542/2.0542 & $-$3.2294/3.2292 & 91.8/77.4   & $-$367/252 &  224.1/254.6\\
C$_{13b}$ & $-$2.0855/2.0855 & $-$3.2254/3.2257 & 92.5/78.2    & $+$369/255 &  224.1/254.6\\
\hline
\end{tabular}
\caption{Recovered image parameters (see Table 1). The flux ratios and 
magnifications (assuming point-source structures) are given by $r_{\rm
\{B,C\}/A}$ and $\mu$, respectively, and the time-delays are given by
$\Delta t$ with respect to the leading image A$_1$ (assuming the
cosmological model mentioned in Sect.1).  The first and second values
indicate those for $\sigma_{\rm M1}$=560 and 175\,km\,s$^{-1}$,
respectively.  The recovered source positions for $\{$A$_{1}$,
B$_{1}$$\}$ and $\{$A$_{2}$, B$_{2}$, C$_{12b}$, C$_{13b}$$\}$ are
(1.214/5.010$''$, 10.143/9.670$''$) and (1.220/5.003$''$,
10.144/9.669$''$), respectively. Even though, the change in source
position is large, when changing $\sigma_{\rm M1}$, this
is no relevance because it not an observable. }
\end{table*}

Finally, we note the remarkably high magnification ($\mu$$\sim$300) at
the brightness peaks (i.e. not integrated over the image) of images
C$_{12}$ and C$_{13}$, although the precise value is sensitive to the
details of the model input (see also below). For example, the
magnification changes by about $\sim$10\% when changing the slope of
the radial mass profile by $\sim$5\%. Even so, this is the highest
(inferred) magnification for any known lens system.  Because it is
primarily directed tangentially and the jet direction is nearly
perpendicular to the fold caustic, according to the model, any
relativistic motion in the radio jet structure will be enhanced by a
factor $\sim$$\mu$. Superluminal velocities of say $\sim$3$h^{-1}\,c$
(e.g. Vermeulen \& Cohen 1994) could therefore lead to {\sl
hyperluminal} velocities of $\sim$10$^3\,h^{-1}\,c$. Similarly, one
can probe structure on intrinsic scales of a few micro-arcsecond when
observed with VLBI. On the other hand, if the lensed structure is part
of the counter-jet (see Sect.3), such high velocities are no longer
expected (even though velocities of $\sim0.5\,h^{-1}\,c$ for the
counter-jet are still likely and could lead to {\sl hyperluminal}
velocities of $\sim10^2\,h^{-1}\,c$).  The absence of strong
variability in the source (Haarsma et al. 2001) and its GPS-type radio
spectrum might, however, be indicative of the absence of strongly
relativistic motion.

One problem that we have not yet addressed is the considerable
difference in angular distance between the pair components
C$_{11}$--C$_{12}$ and C$_{13}$--C$_{2}$, which we did not include in
the lens model.  Given the high magnification and symmetry around the
critical curve, one would expect these distances to be similar. The
fact that they are not seems to argue against our model and in favor
of a two lens model (Nair \& Garrett 1997). However, magnifications
and the magnification matrix are a function of differences in
higher-order derivatives of the local lens potential (see Schneider et
al. 1992). Especially in regions of very high magnification, many of
these derivatives are very close to zero, because in those cases the
lensed images form at very shallow extrema of the time-delay surface
(e.g. Blandford \& Hogg 1996).
It is therefore not inconceivable that even a miniscule perturbation
of the local lens potential (e.g. by a globular cluster, halo
substructure, etc) will have an enormous effect on the local
magnification matrix, i.e. on the image magnifications and the image
positions.  These discrepancies near critical curves have been seen in
other lens systems with very high image magnifications (Mao \&
Schneider 1998) and might not be uncommon in general. As an example,
if we include a typical globular cluster with a velocity dispersion
$\sigma$=7\,km\,s$^{-1}$ (Einstein radius of $\la$1~mas) about 5~mas
away from image C$_2$, it changes the magnifications of C$_2$ and
C$_{11}$ considerably, but also projects them on the same position in
the source plane.  The magnification ratio between C$_{12}$ and
C$_{13}$ remains equal to unity within a few percent. The probability
of such minor perturbations is considerable, especially because one
expects hundreds if not thousands of globular clusters, dwarf
satellites, etc.  around these massive elliptical galaxies. At higher
redshift, according to cosmological CDM models, the amount of halo
substructure could be even more prevalent. We are therefore not too
worried about this apparent discrepancy, but are warned that even
though the magnifications in complex C are very high, their precise
values are quite uncertain and should only be taken as indicative.

\section{Discussion \& Conclusion}

We have shown that 2016+112 can be explained by a {\sl single-screen}
mass model. The observational constraints are reasonably well
reproduced by the proposed model.  The axial ratio of the surface MD
of the lens galaxy~D is somewhat larger than the axial ratio of its SB
distribution, as determined from HST NICMOS {\sl F160W}--band
observations. Also its mass-to-light ratio is in good agreement with
that of other lens galaxies. Surprisingly, our model is consistent
with the presence of a massive component about 1$'$ north--west
of the lens galaxy, suggested from a weak-lensing analysis of the
field (Clowe et al. 2001). We find evidence for an external shear that
might result from a `filamentary structure' of high-redshift galaxies
running across the lens system roughly from east to west (see Soucail
et al. 2001). This filament is not a dynamically related structure,
because it contains galaxies over a wide range of redshifts.  The
massive dark cluster previously suggested can not be confirmed or
excluded, based on the strong-lensing constraints. In particular, a
single $\sim L_*$ galaxy several arcsec east of galaxy D (see Fig.3)
could also be consistent with the strong lensing constraints. A faint
object at that approximate position does appear on a deep R--band
image, i.e. Fig.4 in Clowe et al. (2001), although we do not know
whether that object is indeed a galaxy and what redshift is has
(i.e. whether it forms a second lens screen). 
Given the high redshift of the
source, a perturbing galaxy at that position could have a wide range
of redshifts (0.5--1.5), not necessarily being that of the primary
lens galaxy.

Because we can explain the available data on 2016+112 with a
single-screen model, we conclude that a second screen is at most only
a pertrubation and that the optical/IR object C$'$ is a highly
magnified arc of the AGN host galaxy at $z$=3.273. Thus, we predict
that optical spectroscopy of complex C$'$ will not yield a redshift in
between that of lens galaxy D and the source as previously suspected,
but will instead {\sl only} show features at the source redshift.

Our model predicts a very high magnification ($\mu$$\sim$300) at the
brightness peaks of images C$_{12}$ and C$_{13}$. This can be expected
based on the proximity of these images (only $\sim$20\,mas; see Fig.1) to the
critical curve passing in between them. Because, the
magnification is inversely proportional to the separation between the
two images (e.g. Schneider et al. 1992, Chapt.6), the small separation
($\sim$40\,mas) between the images results in an order of magnitude
higher magnification than normally observed. This is enhanced by the
small angle between the line joining the images and the caustic,
resulting in an even smaller distance of the images to the caustic
(i.e. a higher magnification). A magnification that is 1--2 orders of
magnitude larger than normally seen for two merging images
($\mu$$\sim$10), can therefore be expected. This magnification could
lead to observable {\sl hyperluminal} motion with velocities of order
$\sim$$10^3\,$$h^{-1}\,c$ of micro-arsec-scale structure in the lensed
jet, although the GPS-type radio spectrum and low variability of the
source at present do not support the notion of high apparent
velocities in the radio jet, especially also because our model
suggests that it is the counter-jet that is being magnified.

Our source model appears to be that of a type--II quasar (e.g.  Yamada
et al. 2001; Chartas et al. 2001). At radio wavelengths, only the
counter-jet of the source is quadruply imaged, whereas the AGN core
and radio jet are doubly imaged. The host galaxy and the extended
narrow-line emission around the core is part doubly and part quadruply
imaged, which explains its offset from the optical and radio
structures and also the difference in line ratios between images A \&
B and complex~C (Yamada et al. 2001). The absence of BLR emission and
the presence of X-ray emission (Chartas et al. 2001) further support
the identification of this source as a type--II quasar.

Finally, we suggest that the two patches of Ly--$\alpha$ emission
found by Schneider et al. (1986, 1987) could be cold gas clouds in the
IGM -- possibly around some nearby galaxies -- that are illuminated by
the emission cone coming from the AGN along the jet axes. There are
indeed several objects at the positions of these patches (Soucail et
al. 2001; Clowe et al. 2001), of which the brightest has the same
redshift as the source (Soucail et al. 2001).  When projecting the two
patches on the source plane (Fig.3), using the mass model presented in
this paper, we find that one lies close to the (counter) jet axis in
the source plane, i.e.  $\sim$20$^\circ$, whereas the second patch
makes an angle of $\sim$50$^\circ$. Their distances to the AGN in the
source plane are 1.9 and 1.5 arcsec, respectively. They are therefore
not multiply imaged.  If the jet is pointed towards (or away from) the
observer, the real angles between the patch, AGN and observer could be
smaller. The low expected metalicity for such IGM clouds could explain
the presence of strong Ly--$\alpha$ emission lines and the absence of
strong metal lines, such as the C{\footnotesize IV} lines seen near
images A, B and C$'$ (Schneider et al. 1987; Lawrence 1996). If these
patches are indeed illuminated by the AGN, we expect this Ly--$\alpha$
emission to be highly polarized. Polarimetry observations on a 10-m
class telescope could confirm this.

The ultimate test between the different one and two-lens models will
be the measurement of the redshift of spectral (emission and absorption)
lines from complex C$'$, which can only be
$z\approx$3.273, according to the model proposed in this paper.

\section*{Acknowledgments}

LVEK and RDB would like to thank Ian Browne for invaluable help during the initial
phases of this research. We thank Ian Browne and Sunita Nair for useful comments
on a draft of the paper. LVEK thanks Eric Agol for stimulating discussions and
George Chartas for sending a version of 
their paper on {\sl Chandra} X-ray observations of 2016+112 prior to publication.
This research has been supported by NSF~AST--9900866 and STScI~GO--06543.03--95A.

{}

\end{document}